\documentclass[english,prl,aps,showpacs,twocolumn]{revtex4-1}

\def\be{\begin{equation}}
\def\ee{\end{equation}}
\usepackage{graphicx}
\usepackage{bm}
\usepackage[T1]{fontenc}
\usepackage[latin9]{inputenc}
\usepackage{amsmath}
\usepackage{amssymb}
\usepackage{float}
\usepackage[bookmarks=true,
   colorlinks=true,
   linkcolor=blue,
   urlcolor=blue,
   citecolor=blue,
   bookmarks=true,
   hyperindex=true
]{hyperref}

\begin{document}
\title{Observation of phonon parametric down-conversion in a spherical Bose-Einstein condensate}

\author{Tianyou Gao$^{1,6}$, Jian-Song Pan$^{3}$, Dongfang Zhang$^{1}$, Lingran Kong$^{1,6}$, Ruizong Li$^{1,6}$, Xing Shen$^{1,6}$, Xiaolong Chen$^{4}$, Shi-Guo Peng$^{1}$, Mingsheng Zhan$^{1,5}$}

\author{W. Vincent Liu$^{2,3,5}$}
\email{wvliu@pitt.edu}

\author{Kaijun Jiang$^{1,5}$}
\email{kjjiang@wipm.ac.cn}

\affiliation{$^{1}$State Key Laboratory of Magnetic Resonance and Atomic and Molecular Physics, Wuhan Institute of Physics and Mathematics, Chinese Academy of Sciences, Wuhan, 430071, China}
\affiliation{$^{2}$Department of Physics and Astronomy, University of Pittsburgh, Pittsburgh, Pennsylvania 15260, USA}
\affiliation{$^{3}$Wilczek Quantum Center, School of Physics and Astronomy and T. D. Lee Institute, Shanghai Jiao Tong University, Shanghai 200240, China}
\affiliation{$^{4}$Centre for Quantum and Optical Science, Swinburne University of Technology, Melbourne 3122, Australia}

\affiliation{$^{5}$Center for Cold Atom Physics, Chinese Academy of Sciences, Wuhan, 430071, China}
\affiliation{$^{6}$School of Physics, University of Chinese Academy of Sciences, Beijing 100049, China}

\date{\today}

\begin{abstract}
We report the observation of parametric down-conversion of phonons in a spherical Bose-Einstein condensate. The spherical symmetry, which is crucial for observing this phenomenon, is experimentally demonstrated by measuring the collective mode and expansion behavior of the condensate. The low-energy monopole mode is excited by coupling with a high-energy mode with a nearly twice eigen-frequency. The population of the low-energy mode becomes maximum only when the high-energy mode is resonantly excited. Furthermore, we directly observe the parametric down-conversion process in the driving process, through simultaneously probing the two coupling modes. The experimental observation is consistent with the perturbation theory including the gravity effect. This work opens the challenge in related study of the condensate beyond mean-field theory and has potential applications in quantum information.
\end{abstract}

\maketitle

\emph{Introduction}.--
Spontaneous parametric down-conversion (SPDC) of photons in nonlinear crystal has been widely used to produce high-qualified correlated photon pairs in quantum techniques~\cite{Pershan1962PR, Tang1968PRparametric, Tang1968PR, Pan2012multiphoton, Pan2013Nonlocal}. In SPDC process, one high-frequency photon is split as two low-frequency photons, in accordance with the conservation laws of energy and momentum. The nonlinear third-order process and quantum fluctuation are responsible for the SPDC phenomenon. It is of fundamental interest to simulate optical phenomena with matter wave since the observation of Bose-Einstein condensate (BEC)~\cite{Andrews1997Observation, Inouye1999Phase, Phillips1999Nature}. The dynamics of BEC can be approximately described with the nonlinear Gross-Pitaevskii equation. Hence it is naturally asked if similar phenomenon like SPDC exists for the Bogoliubov collective modes (called phonons) in a BEC, where the theory beyond mean-field approximation is required. While a nonlinear coupling between a high-energy mode and a low-energy mode has been reported in an anisotropic BEC~\cite{Foot2000PRL, Foot2001PRL}, the possibility that the observed phenomenon is a SPDC process is excluded when considering the parity symmetry~\cite{Foot2002PRA, StoofPRA2001}. Experimentally observing the phonon parametric down-conversion (PPDC) in quantum gases is not confirmed up to now.

In previous experiments on quantum gases performed mostly in anisotropic traps with less or no well-defined symmetry~\cite{Wieman1996PRL, Ketterle1996PRL, Dalibard2002PRL, Thomas2004PRL, Grimm2007PRL, Salomon2009PRL, Grimm2013PRA, Grimm2013PRL}. the dense excitation spectra make it difficult to observe the specific mode-coupling process. To clearly observe the phonon version of SPDC, the better choice is to prepare a BEC with a good sphericity. In a spherical BEC, the collective modes are well characterized with different spatial symmetries, which makes it easy to excite a specific mode. Furthermore, the symmetries of the wave functions and the conservation laws strictly limit the allowed coupling between collective modes, which facilitates definitely observing a specific mode-coupling process.

In this letter, we report the first experimental observation of the PPDC in quantum gases. In order to observe this coupling process, we produce a spherical rubidium BEC in a optical dipole trap. The isotropy of the BEC is proved by the measurement of the expansion behavior as well as the oscillation of the quadrupole mode. When exciting a high-energy mode, we find the low-energy monopole mode with a nearly half eigen-frequency is produced. The oscillation amplitude of the low-energy mode becomes maximum only when the high-energy mode is resonantly excited, which demonstrates that the low-energy mode originates from coupling with the high-energy mode. We further directly probe the high-energy and low-energy mode during the driving process and find that the PPDC process occurs after some driving time (about 21 ms). The perturbation theory including the gravity effect explains the permissibility of the coupling between these two modes.

\begin{figure}
\centerline{\includegraphics[width=8cm]{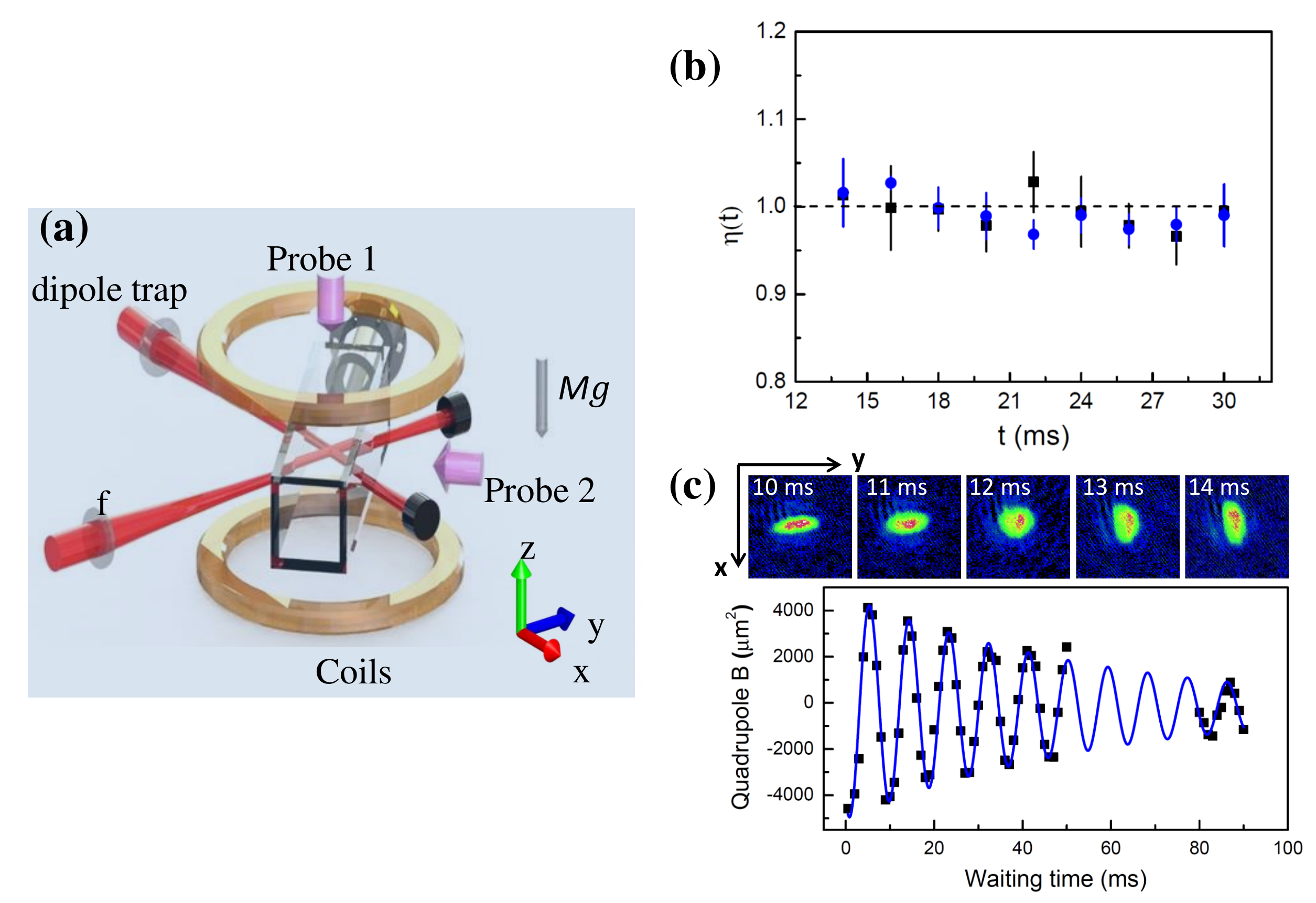}}
\caption{(color online) Creating a spherical Rb BEC. (a) Experimental setup. The optical dipole trap is composed of two focused red-detuned laser beams along $x$ and $y$ directions. The gravity is along the $-z$ direction. Ultracold atoms are simultaneously probed in the vertical and horizontal directions. (b) Aspect ratio $\eta(t)$ versus the free expansion time t. The black squares (blue circles) are for the images probed in the horizontal (vertical) direction. The error bars indicate the uncertainties for three measurements. (c) Oscillation of the quadrupole mode. The upper row shows atom clouds for five waiting times in the trap. The lower row shows the oscillation of the parameter $B=R_{x}^2-R_{y}^2$. The blue solid curve is the numerical fitting with a damped sinusoidal function. }  \label{Fig1}
\end{figure}

\begin{figure*}
\centerline{\includegraphics[width=16cm]{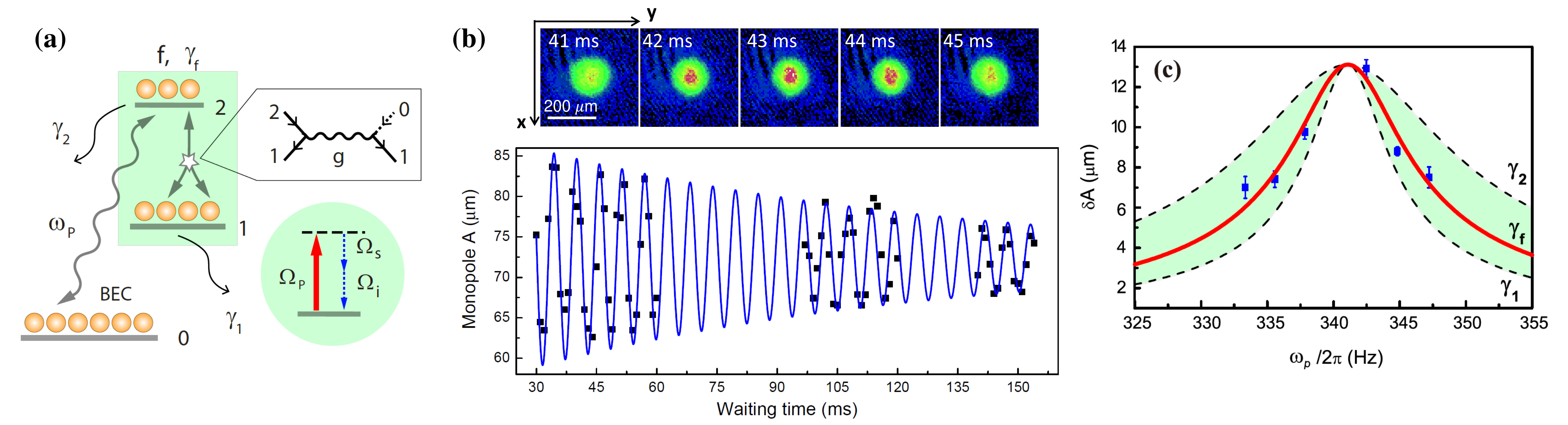}}
\caption{(Color online) Observing the phonon parametric down-conversion (PPDC). (a) Schematics for the PPDC. Bose condensate (mode $0$) is excited with a frequency $\omega_p$ to the mode $2$ ($n=2, l=1$). Mode $2$ and mode $1$ (monopole mode) are coupled by the interaction. As illustrated in the open pentagon, one phonon 2 and one phonon 0 merge as two phonons 1. The coupled two modes can be considered as one effective mode $f$. $\gamma_{i=1, 2, f}$ denote the Landau damping rates of the mode $i$. As a comparison, the PDC of photons is shown in the rounded shadow box. (b) Probing the low-energy monopole mode. The upper row shows atom clouds for five waiting times in the trap. The lower row shows the oscillation of the effective width $A=\sqrt{R_{x}^2+R_{y}^2+R_{z}^2}$. The blue solid curve is the numerical fitting with a damped sinusoidal function. (c) Oscillation amplitude $\delta A$ of the monopole mode versus $\omega_p$. The error bars denote the uncertainties in fitting the oscillation as in (b). Two dashed curves are the Lorentz distributions with widths of $\gamma_{1}=2\pi\times 1.42$ Hz and $\gamma_{2}= 2\pi\times 3.23$ Hz, given by the numerical calculation. The red solid curve is the numerical fitting of the experimental data with a Lorentz function, giving $\gamma_{f} = 2\pi\times 2.01$ Hz. The heights are normalized by the experimental data. }
\label{Fig3}
\end{figure*}

\emph{Preparing spherical BEC}.--
We produce a spherical $^{87}$Rb BEC in an optical dipole trap in which the trapping frequencies along $x,y,z-$directions are the same. The experimental setup is shown in Fig.~\ref{Fig1}(a), which is developed from the Ref. \cite{Jiang2016CPL}. The spherical trap is composed of the optical dipole trap and the gravity. The combined trap is given by
\begin{equation} \label{eq:trap}
\begin{split}
U \left(x,y,z\right) =& -U_{1}\exp\left(-\frac{2x^{2}}{w_{1x}^{2}}-\frac{2z^{2}}{w_{1z}^{2}}\right)\\
&-U_{2}\exp\left(-\frac{2y^{2}}{w_{2y}^{2}}-\frac{2z^{2}}{w_{2z}^{2}}\right)-Mgz,
\end{split}
\end{equation}
where $w_{1x}$ ($w_{2y}$) and $w_{1z}$ ($w_{2z}$) are the waists of the optical trap beam along the $y$ ($x$) direction. With aid of the gravity force $Mg$, the condition to form a spherical trap becomes $U_1/w_{1x}^2=U_2/w_{2y}^2$, where $U_1$ and $U_2$ are the peak potentials of the two beams, respectively. In this case, we can accurately adjust the relative intensities of the two beams to form a spherical BEC. By measuring the oscillation of the center of the mass (COM) of the atomic cloud in the trap, we get the mean trapping frequency $\bar{\omega}= \left(\omega_{x}+\omega_{y}+\omega_{z}\right)/3 = 2\pi\times 77.5$ Hz. The asphericity $A=(\omega_{max}-\omega_{min})/\bar{\omega}\approx3.7\%$, where $\omega_{max}$, $\omega_{min}$ are the maximum and minimum trapping frequencies along three directions, respectively. Whenever measuring collective modes, the trapping frequencies are always calibrated to keep the asphericity $A <5\%$. We improve the position stability of the optical trap beam better than 3 $\mu$m to achieve a stable spherical BEC (see Supplemental Material for technical details \cite{Supplemental}). The atoms stay in the spin state $|F, m_{F}\rangle = |1, -1\rangle$. The atomic temperature is about 80 nK from analyzing the free expansion of the atomic cloud, and the atom number of BEC is about $1.2\times10^5$.

The aspect ratio $\eta(t)$ of the condensate during the free expansion is probed in Fig.~\ref{Fig1}(b) by suddenly switching off the optical trap. For the images probed along the horizontal direction, $\eta(t)=R_{\parallel}(t)/R_z(t)$ where $R_{\parallel}(t)$ and $R_z(t)$ are the Thomas-Fermi radius in the horizontal and vertical directions, respectively. For the images probed from the vertical direction, $\eta(t)=R_{x}(t)/R_y(t)$ where $R_{x}(t)$ and $R_y(t)$ are the Thomas-Fermi radius in the $x$ and $y$ directions, respectively. $\eta(t)$ remains unity during the free expansion, which is unique for a spherical BEC. For an anisotropic BEC, the expansion is anisotropic and the aspect ratio $\eta(t)$ approaches an asymptotic value dependent on the ratio of the trapping frequencies \cite{Rempe1998EPL, Ketterle1998PRLexpansion, Stringari1999RMP}.

In a spherical BEC, the collective mode spectrum is simplified as $\omega(n,l)=\omega_{0}(2n^{2}+2nl+3n+l)^{1/2}$ where $l$ is the angular momentum number, $n$ is the principle number and $\omega_{0}$ is the trapping frequency~\cite{Stringari1996PRL}. To confirm the analytical calculation of the collective mode based on the spherical-trapping assumption, we excite the quadrupole mode ($n=0, l=2$) according to its symmetry with the eigen-frequency. We modulate the intensity of the optical trap beam along the $x$ direction with a frequency of about $\sqrt{2} \omega_0$. The modulation lasts ten periods and then the condensate is probed for different waiting time in the trap, using the absorption-imaging method with a time of flight (TOF) of 28 ms. The condensate is compressed and decompressed simultaneously in $y$ and $z$ directions, and then it oscillates out-of-phase between the $y-z$ plane and the $x$ direction. The modulation amplitude $\delta \omega_{y0} /\omega_{y0}$ along $y$ direction is about 12$\%$, which is in the linear response regime. The images probed from the vertical direction are used to analyze the relative motion between $x$ and $y$ directions (see Fig.~\ref{Fig1}(c)). We define a parameter $B=R_{x}^2-R_{y}^2$ to quantitatively describe the quadrupole mode, where $R_{i} (i=x,y)$ is the Thomas-Fermi radius of the condensate. As shown in Fig.~\ref{Fig1}(c), the experimental data are fitted using a damped sinusoidal function $B(t)=B_0+\delta B \exp(-t/t_0) \sin (\omega_Q t+\phi)$. From the fitting, $\omega_Q = 2\pi\times111.21(32)$ Hz $=1.435(4)\omega_0$, which is consistent with the quadrupole mode eigen-frequency ($\sqrt{2} \omega_0$) for the spherical Bose condensate \cite{Stringari1996PRL}. The lifetime of the quadrupole mode is $t_0 = 58.2(66)$ ms, which is also consistent with the calculated value $49.3$ ms of the Landau damping in the spherical condensate (see Supplemental Material for the calculation details \cite{Supplemental}). The statistics errors come from the uncertainty in the fitting process.

\emph{Theoretical description of PPDC process}.--
Expanding the interaction term with respect to the fluctuation above the ground state up to the second-order is the central idea to calculate collective modes in the Bogoliubov theory~\cite{Bogolyubov1947On}. The third-order expansion gives rise to the coupling between the collective modes. In an ideal spherical trapping potential, it is hard to find the SPDC-type coupling between the collective modes, which is required to satisfy the matching condition at the same time.  While here the gravitational force makes the isotropic potential slightly deformed and develops an odd parity in the z-direction. Seeing the Eq.~(\ref{eq:trap}), the first-order term of the trap is absorbed by the shift of the trap center, then we only consider the third-order deformation $V^{'}=\lambda z^3$ of the trap. The deformation is weak and a perturbation analysis is applicable~(see Supplemental Material \cite{Supplemental}). We find mode ($n=2, l=1$) not only can be coupled to the monopole mode ($n=1, l=0$) through a mechanism like SPDC, but also have an eigen-frequency nearly twice of the latter. For convenience, we call the monopole mode as mode $1$, the mode ($n=2, l=1$) as mode $2$ and the stationary Bose condensate as mode $0$. The coefficient characterizing the probability of the down-conversion process is given by \cite{Morgan1998nonlinear, Foot2002PRA}
\begin{equation}\label{eq:M12}
\begin{split}
M_{12}=&2\int d\boldsymbol{r}\psi_{0}^{'}[\left(2\tilde{u}_{1}^{\ast}\tilde{\upsilon}_{1}^{\ast}+\tilde{u}_{1}^{\ast}\tilde{u}_{1}^{\ast}\right)\tilde{u}_{2}\\
&+\left(2\tilde{u}_{1}^{\ast}\tilde{\upsilon}_{1}^{\ast}+\tilde{\upsilon}_{1}^{\ast}\tilde{\upsilon}_{1}^{\ast}\right)\tilde{\upsilon}_{2}],
\end{split}
\end{equation}
where $\left(\begin{array}{cc}
\tilde{u}_{\nu} & \tilde{\upsilon}_{\nu}\end{array}\right)=\left(\begin{array}{cc}
u_{\nu}^{'} & \upsilon_{\nu}^{'}\end{array}\right)-c_{\nu}\left(\begin{array}{cc}
\psi_{0}^{'} & -\psi_{0}^{'\ast}\end{array}\right)$ with $c_{\nu}=\int d\boldsymbol{r}\psi_{0}^{'}u_{\nu}^{'}\approx-\int d\boldsymbol{r}\psi_{0}^{'}\upsilon_{\nu}^{'}$ is the orthogonalized collective mode wave function with respect to the ground-state wave function. The prime denotes the normalized perturbed wave functions.

The PPDC process is described by the Lagrangian $\mathcal{L}=i\hbar\sum_{j=1,2}\hat{b}_{j}^{\dagger}\partial_{t}\hat{b}_{j}+\hbar\kappa(\hat{b}_{2}^{\dagger}\hat{b}_{1}^{2}e^{i\Delta t}+\hat{b}_{1}^{\dagger2}\hat{b}_{2}e^{-i\Delta t})/2$, where $\hat{b}_j$ ($\hat{b}_j^\dagger$) annihilates (creates) a phonon of mode $j$,  $\kappa=|NgM_{12}/\hbar|$ is the coupling coefficient, and $\Delta=(\omega_2^{'}-2\omega_1^{'})$ with the perturbed eigen-frequency $\omega_j^{'}$ is the detuning \cite{Foot2002PRA}. In our experiment, $\lambda\approx0.0175 a_H^{-3}$, $\Delta\approx0.2\omega= 2\pi\times 15.50$ Hz and $\kappa \approx 0.90\omega_{0} = 2\pi \times 69.75$ Hz. The effective coupling strength for the PPDC process can be estimated as $\xi=|\beta_{2}\kappa|^2/\Delta \approx 3.98|\beta_{2}|^2\omega_{0} = 2\pi \times 308.45|\beta_{2}|^2$, where $\beta_{2}=\langle\hat{b}_2\rangle$ is the mean-field probability amplitude of mode 2. The non-zero $M_{12}$ indicates the PPDC process is permissible in our experiment. Although the mean-field treatment of the Euler-Lagrange equations derived from $\mathcal{L}$ gives the parametric up-conversion process when only mode 1 is initially occupied \cite{Foot2000PRL, Foot2002PRA}, it can not give the down-conversion process when only mode 2 is initially occupied~\cite{Foot2002PRA, StoofPRA2001}. The method beyond mean-field theory is desired to quantitatively describe the parametric down-conversion process.

\emph{Observation of PPDC process}.--
We drive the high-energy mode ($n=2, l=1$) by periodically modulating the trapping potential with a frequency $\omega_{p}\approx 2\pi\times 340.00$ which is nearly equivalent to its eigen-frequency $\sqrt{19}\omega_0$. Mode ($n=2, l=1$) corresponds to the oscillation of the center-of-mass of the condensate. The intensity variations of the two beams are accurately balanced so that the condensate keeps nearly a sphere when oscillating in the trap. In our trapping configuration of Eq.~(\ref{eq:trap}), the potential minimum along the $z$ direction is given by $z_{0} = \frac{Mg}{4(U_{1}/w_{1z}^{2}+U_{2}/w_{2z}^{2})}$. So the modulation will makes $z_{0}$ oscillate with frequency $\omega_{p}$. This oscillation corresponds to the mode ($n=2, l=1, m=0$)~(see Supplemental Material \cite{Supplemental}), where $m$ is the axial projection of the angular moment. After the modulation lasts about ten periods, we find the condensate widths in $x,y,z$-directions oscillate synchronously in the trap (see Fig.~\ref{Fig3}(b)). The in-phase oscillation in three directions indicates the monopole mode is indeed excited. We define an effective width $A=\sqrt{R_{x}^2+R_{y}^2+R_{z}^2}$ to quantitatively characterize the monopole oscillation. In Fig.~\ref{Fig3}(b), we numerically fit the experimental data using a damped sinusoidal function $A(t)=A_0+\delta A \exp(-t/t_0) \sin (\omega_M t+\phi)$. From the fitting, $\omega_M = 2\pi\times176.93(31)$ Hz $=2.283(4)\omega_0$, which is consistent with the monopole mode eigen-frequency $\sqrt{5} \omega_0$ \cite{Stringari1996PRL}. The lifetime of the monopole mode is $t_0 = 104.4(57)$ ms, which is also consistent with the numerically calculated value 111.9 ms of Landau damping~(see Supplemental Material \cite{Supplemental}).

\begin{figure}
\centerline{\includegraphics[width=7cm]{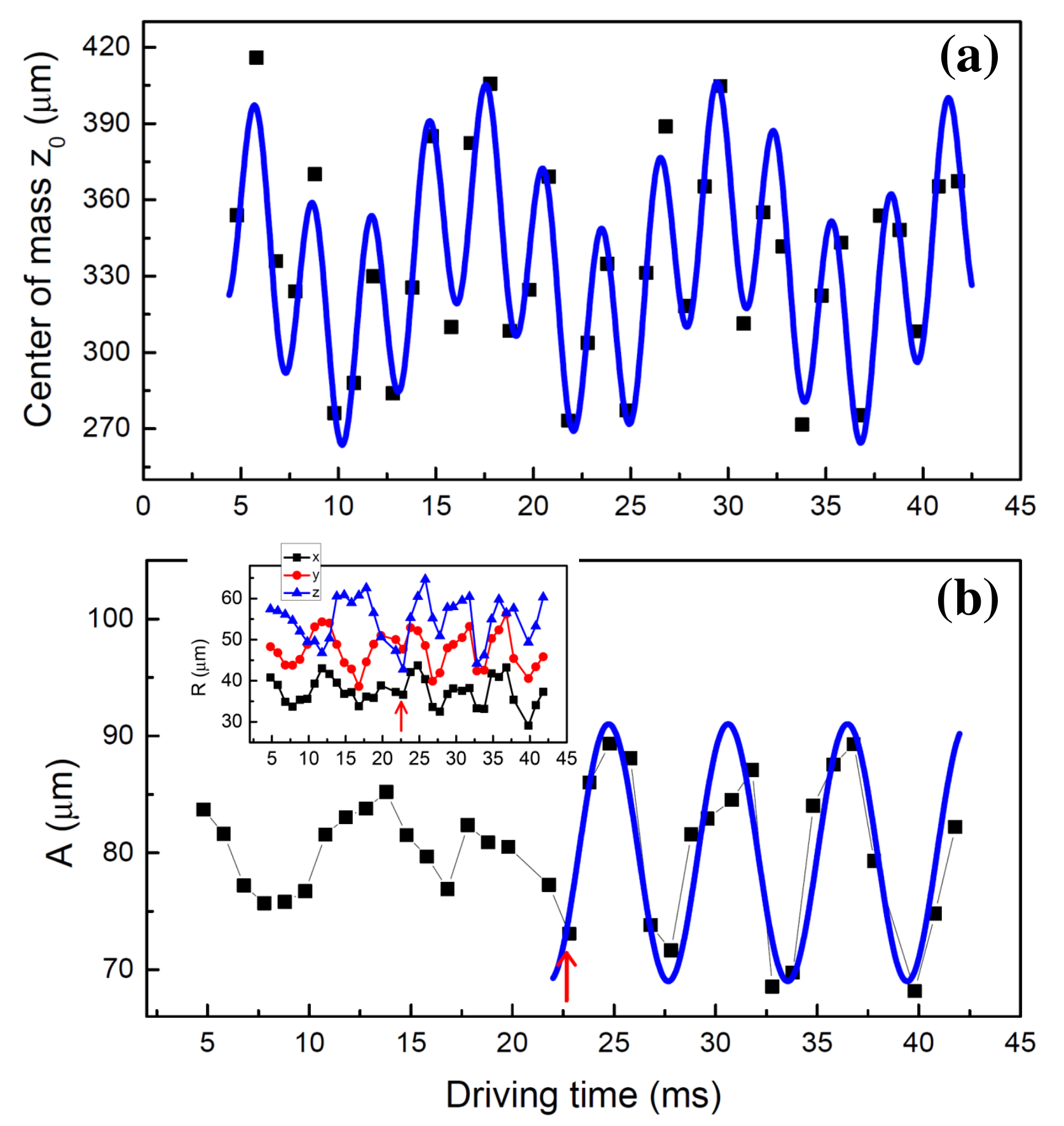}}
\caption{(Color online) Directly probing the phonon parametric down-conversion (PPDC) in the driving process. (a) The center of mass of the atomic cloud along the z direction versus the driving time. The experimental data are fitted with a double-sine function. The obtained two oscillation frequencies, 1.01$\omega_{0}$ and 4.36$\omega_{0}$, correspond to the dipole mode and the mode ($n=2, l=1$), respectively. (b) Observation of the monopole mode. When the driving time is larger than 21 ms (indicated by the red arrow), the condensate widths in $x, y, z$-directions begin to oscillate in-phase (see the inset), indicating the monopole mode. Fitting the effective width$A=\sqrt{R_{x}^2+R_{y}^2+R_{z}^2}$ with a sinusoidal function gives the oscillation frequency $\omega = 2\pi\times170.13(173)$ Hz $ \approx 2.20(2)\omega_{0}$. }\label{Fig4}
\end{figure}

The driving process of mode ($n=2, l=1$) (mode 2) and the coupling mechanism between mode 2 and mode 1 (monopole mode) are illustrated in Fig.~\ref{Fig3}(a). Since modes $1$ and $2$ can couple back and forth and the total energy is conserved, the two modes can be considered as an effective mode $f$. The atom population transferred from the ground state $0$ to state $f$ is proportional to the total excitation energy in the driving process. The width of the population distribution with respect to the driving frequency is proportional to the effective damping rate $\gamma_f$ of state $f$. The total excitation energy finally flows to mode $1$ through the PPDC process in the long time limit, and can be monitored by measuring the amplitude of the monopole oscillation. Under this picture, $\gamma_f$ locals in the range $[\gamma_1,\gamma_2]$, where $\gamma_{1}$ and $\gamma_{2}$ are the Landau damping rates of mode $1$ and mode $2$, respectively. The specific value of $\gamma_f$ depends on uncontrollable details like initial conditions. As shown in Fig.~\ref{Fig3}(c), the observed distribution of the amplitude of the monopole oscillation well supports the above picture. The amplitude of the monopole mode is very sensitive on the driving frequency and becomes maximum only when the mode ($n=2, l=1$) is resonantly excited. This demonstrates that the oscillation of the monopole mode originates from the mode ($n=2, l=1$) through the SPDC process. Unlike the SPDC of photons, where a pump phonon ($\Omega_p$) is split as a signal photon ($\Omega_{s}$) and an idle photon ($\Omega_{i}$) via the optical nonlinear coupling (in general, the maximum conversion possibility locates at $\Omega_{s}=\Omega_{i}$), one high-energy phonon (mode 2) is split as two low-energy phonons (mode 1) by colliding with the ground-state atoms (mode 0) (see Fig.~\ref{Fig3}(a)).

In Fig.~\ref{Fig4}, we directly observe the parametric down-conversion from the mode ($n=2, l=1$) to the monopole mode during the mode-excitation process. Here we modulate the optical dipole trap for different time and then probe the position and width of the condensate without waiting time in the trap. For mode ($n=2, l=1$), only the center-of-mass of the atomic cloud oscillates while the size does not change. This facilitate directly probing the two modes individually. The oscillation of the center of mass along the $z$ direction is shown in Fig.~\ref{Fig4}(a). We use a double-sine function $z_{0}(t)=z_{0}^{'}+z_{01}\sin(\omega_{1}(t-t_{1}))+z_{02}\sin(\omega_{1}(t-t_{2})/\sqrt{19})$ to fit the experimental data. $\omega_{1} = 2\pi\times338.07(91)$ Hz $= 4.36(1)\omega_{0}$ is very close to the eigen-frequency of the mode ($n=2, l=1$), and $\omega_{1}/\sqrt{19}$ is close to the eigen-frequency of the dipole mode ($n=0, l=1$). $z_{01}^{2}/z_{02}^{2}=2.15$ means that the population in mode ($n=2, l=1$) is about twice that of the dipole mode. This is reasonable because we resonantly excite the mode ($n=2, l=1$) while the dipole mode is excited with a far detuning. We also simultaneously probe the atomic size in Fig.~\ref{Fig4}(b). The atomic widths in $x, y, z$-directions change without a constant phase for a short driving time, but start to oscillate in-phase when the driving time is longer than 21 ms. This indicates that the monopole mode is produced after certain driving time. We fit the effective width $A=\sqrt{R_{x}^2+R_{y}^2+R_{z}^2}$ with a sinusoidal function and obtain the oscillation frequency $\omega = 2\pi\times170.13(173)$ Hz $ \approx 2.20(2)\omega_{0}$, which is very close to the monopole mode eigen-frequency $ \sqrt{5}\omega_{0}$.

\emph{conclusion}.--
We observe the PPDC process in quantum gases for the first time. In order to observe this coupling process, we produce a spherical BEC in a optical dipole trap by overcoming previous technical challenges. Considering the enhancement of the trapping potential, the PPDC observed here resembles more like the cavity-enhanced SPDC \cite{Ou1999cavity} than that in free space. The Beliaev damping of quasiparicle in the continuous spectrum regime has been observed \cite{Davidson2002PRL}. In that case, the mode coupling needs to integrate various scattering modes. In the PPDC process, one high-energy collective mode is split as two low-energy modes with a half eigen-frequency, where the excitation spectrum is discrete. Our reported observation opens the challenge in related study of the condensate beyond mean-field theory. Also, the interaction between phonons in ultracold quantum gases can be tuned by using a magnetic or optical field, which makes the PPDC phenomenon has potential applications in quantum information.

We acknowledge fruitful discussions with Gora Shlyapnikov, David Papoular, and Shizhong Zhang. This work has been supported by the NKRDP (National Key Research and Development Program) under Grant No. 2016YFA0301503, NSFC (Grant No. 11474315, 11674358, 11434015) and CAS under Grant No. YJKYYQ20170025. J.-S. P. acknowledges support from National Postdoctoral Program for Innovative Talents of China under Grant No. BX201700156.

Tianyou Gao, Jian-Song Pan and Dongfang Zhang contributed equally to this work.

\newpage
\begin{widetext}
\appendix

\section*{SUPPLEMENTARY INFORMATION}

\subsection*{Driving mode ($n=2, l=1$) without destroying the spherical symmetry}

In order to drive mode ($n=2, l=1$) without destroying the spherical symmetry, we simultaneously modulate the intensities of the two trapping beams, i.e. $I_i (t)=I_{i0}+\delta I_{i0} \sin{\omega_{p} t}$ ($i=1, 2$). The modulation frequency of the two beams is about $\sqrt{19}\omega_{0}$ which is the eigen frequency of mode ($n=2, l=1$). $I_{i0}$ is the initial optical intensity. The variations $\delta I_{i0} ($i=1, 2$)$ of the two beams are accurately balanced so that the condensate remains nearly a sphere when oscillating in the trap. In our trapping configuration, the potential minimum along the $z$ direction is given by $z_{0} = \frac{Mg}{4(U_{1}/w_{1z}^{2}+U_{2}/w_{2z}^{2})}$. Changing the optical intensities excites the center of mass along the $z$ direction, such that the mode ($n=2, l=1, m=0$) is excited.

\subsection*{Manipulating the optical dipole trap to form a stable spherical BEC}

Forming a good spherical BEC requires that the positions of two optical trap beams can be adjusted with a high accuracy and stability. We use the combination of one acousto-optic modulator (AOM) and one PZT-driven mirror to adjust the position of one beam as shown in Fig. \ref{Figure5}. The beam position can be adjusted with an accuracy of about 1 $\mu$m. The temperature shift of the AOM has a big effect on the spatial stability of the laser beam. To solve this problem, we keep the AOM open as long as possible. A flipped mirror is used to switched on and off the laser beam. The AOM is only switched off about 100 ms for probing atoms in the time of flight. In this way the AOM keeps working for most time in the experimental period of 20 seconds and the temperature change is about 0.5 $^{0}$C. The position stability of the trap beam is better than 3 $\mu$m.

\subsection*{Bogoliubov spectrum of a spherical condensate}
The Hamiltonian of our system can be written as
\begin{equation}\label{eq:Ham}
H=\int d^{3}\boldsymbol{r}\hat{\psi}^{\dagger}\left(-\frac{\hbar^{2}}{2M}\nabla^{2}+V\left(\boldsymbol{r}\right)+\frac{g}{2}\hat{\psi}^{\dagger}\hat{\psi}\right)\hat{\psi},
\end{equation}
where $\hat{\psi}$ is the field operator of the bosons, $V\left(\boldsymbol{r}\right)=\frac{1}{2}M\omega^{2}r^{2}$ is the trapping potential with the trapping frequency $\omega$ and $g=4\pi\hbar^2a_s/M$ with the scattering length $a_s$ is the interaction coefficient. In our experiment, $\omega\approx 2\pi \times 77.5$ Hz and the characteristic length of the trapping potential $a_H=\sqrt{\hbar/M\omega}\approx1.24$ $\mu m$. Obviously, this Hamiltonian possesses the $SO(3)$ rotation symmetry and $U(1)$ gauge symmetry.

The dynamics of the BEC can be described by the Gross-Pitaevskii (GP) equation\cite{Gross1961structure, Gross1963hydrodynamics, Pitaevskii1961vortex}
\begin{equation}\label{eq:GPE}
i\hbar\frac{\partial\psi}{\partial t}=\left(-\frac{\hbar^{2}}{2M}\nabla^{2}+V\left(\boldsymbol{r}\right)+g\left|\psi\right|^{2}\right)\psi,
\end{equation}
where $\psi=\langle \hat{\psi} \rangle$ is the condensate wave function.

In order to find the collective modes of the BEC, we rewrite $\psi$ as
\begin{equation}\label{eq:PsiExp}
\psi\left(\boldsymbol{r},t\right)=e^{-i\mu t/\hbar}\left[\psi_{0}\left(\boldsymbol{r}\right)+\sum_{j}\left(u_{j}\left(\boldsymbol{r}\right)e^{-i\omega_{j}t}+\upsilon_{j}^{\ast}\left(\boldsymbol{r}\right)e^{i\omega_{j}t}\right)\right],
\end{equation}
where $\psi_0$ is the ground-state wave function, $\mu$ the chemical potential, $u_j$ and $\upsilon_j$ the "particle" and "hole" components with eigen-frequency $\pm\omega_j$ respectively of the Bogoliubov transformations. Substituting Eq. (\ref{eq:PsiExp}) into Eq. (\ref{eq:GPE}), we derive the Bogoliubov-de Gennes (BdG) equation
\begin{equation}\label{eq:BogoE}
\left(\begin{array}{cc}
-\frac{\hbar^{2}}{2M}\nabla^{2}-\mu+V\left(\boldsymbol{r}\right)+2g\left|\psi_{0}\right|^{2} & g\psi_{0}^{2}\\
-g\psi_{0}^{\ast2} & -\left(-\frac{\hbar^{2}}{2M}\nabla^{2}-\mu+V\left(\boldsymbol{r}\right)+2g\left|\psi_{0}\right|^{2}\right)
\end{array}\right)\left(\begin{array}{c}
u_{j}\\
\upsilon_{j}
\end{array}\right)=\hbar\omega_{j}\left(\begin{array}{c}
u_{j}\\
\upsilon_{j}
\end{array}\right).
\end{equation}
Here the ground-state wave function $\psi_0$ and the chemical potential $\mu$ are determined by the stationary GP equation
\begin{equation}\label{eq:GPE_stationary}
\left(-\frac{\hbar^{2}}{2M}\nabla^{2}+V\left(\boldsymbol{r}\right)+g\left|\psi_0\right|^{2}\right)\psi_0=\mu\psi_0,
\end{equation}
and the particle number equation $N_{0}=\int d^{2}\boldsymbol{r}\left|\psi_{0}\right|^{2}$, where $N_0$ is the ground-state particle number. Under the parameters of our experiment, since the coefficient reflecting the ratio between the interaction energy and the kinetic energy $N_0 a_s/a_H\approx560\gg1$, the Thomas-Fermi approximation is applicable,
\begin{equation}\label{eq:TFA}
\psi_{0}\approx\begin{cases}
\left(\mu-M\omega^{2}r^{2}\right)^{1/2} & ,\quad r\leq\sqrt{\mu/M\omega^{2}}\\
0 & ,\quad r>\sqrt{\mu/M\omega^{2}}
\end{cases}.
\end{equation}

Since Hamiltonian (\ref{eq:Ham}) possesses the rotation symmetry, Eq. (\ref{eq:BogoE}) can be divided into independent sectors denoted by the angular momentum quantum number $(l,m)$, where $l=0,1,2,...$ and $m=-l,-l-1,...l$ are the quantum numbers arising from the total angular momentum and the z-component of the angular momentum respectively. Under the spherical coordinate system $\boldsymbol{r}=(r,\theta,\phi)$, we expand $(u_j,\upsilon_j)$ as
\begin{equation}\label{eq:lmExp}
\left(\begin{array}{c}
u_{j}\left(\boldsymbol{r}\right)\\
\upsilon_{j}\left(\boldsymbol{r}\right)
\end{array}\right)=\sum_{lm}\left(\begin{array}{c}
u_{j}^{\left(lm\right)}\left(r\right)\\
\upsilon_{j}^{\left(lm\right)}\left(r\right)
\end{array}\right)Y_{lm}\left(\theta,\phi\right),
\end{equation}
where $Y_{lm}\left(\theta,\phi\right)$ is the spherical harmonic wave function. Substituting Eq. (\ref{eq:lmExp}) into Eq. (\ref{eq:BogoE}) and integrating the two sides with $\int_{0}^{2\pi}d\phi\int_{0}^{\pi}d\theta Y_{lm}\left(\theta,\phi\right)\sin\theta$, we derive
\begin{equation}\label{eq:lmBogo}
\left(\begin{array}{cc}
h_{0}\left(r\right)+2g\left|\psi_{0}\right|^{2} & g\psi_{0}^{2}\\
-g\psi_{0}^{\ast2} & -\left(h_{0}\left(r\right)+2g\left|\psi_{0}\right|^{2}\right)
\end{array}\right)\left(\begin{array}{c}
u_{j}^{\left(lm\right)}\left(r\right)\\
\upsilon_{j}^{\left(lm\right)}\left(r\right)
\end{array}\right)=\hbar\omega_{j}^{\left(lm\right)}\left(\begin{array}{c}
u_{j}^{\left(lm\right)}\left(r\right)\\
\upsilon_{j}^{\left(lm\right)}\left(r\right)
\end{array}\right),
\end{equation}
where $h_{0}\left(r\right)=-\frac{\hbar^{2}}{2M}\left[\frac{1}{r^{2}}\frac{d}{dr}\left(r^{2}\frac{d}{dr}\right)-\frac{l\left(l+1\right)}{r^{2}}\right]+V(r)$. The above equation is solved numerically. Since the transformation $\{(\begin{array}{cc}
u_{j} & \upsilon_{j}\end{array}),\hbar\omega_{j}\}\longrightarrow\{(\begin{array}{cc}
\upsilon_{j}^{\ast} & u_{j}^{\ast}\end{array}),-\hbar\omega_{j}\}$ doesn't change Eq. (\ref{eq:BogoE}), one can get the wave functions of the "hole" component by looking at the wave function of the "particle" component. For convenience, we sort the non-negative eigen-energy from small to large and denote it with integer $j$ from $0$ to infinity for each sector $(l,m)$. The lowest excited collective mode with spherical symmetry, mode $(n,l)=(1,0)$, is called monopole mode, while the lowest excited mode with $l=1$, mode $(0,1)$ and that with $l=2$, mode $(0,2)$, are respectively called dipole mode and quadrupole mode.

For convenience, we use the Dirac brackets $|\psi_j\rangle$ with the eigen-energy $E_j$ to denote the $j$-th zero-order collective mode (i.e. $\langle\boldsymbol{r}|\psi_{j}\rangle=(\begin{array}{cc}
u_{j} & \upsilon_{j}\end{array})^{T}$, $E_j=\hbar\omega_j$). The zero-energy Goldstone mode $(\psi_{0},-\psi_{0}^{\ast})^{\dagger}$ is denoted with $|\psi_0\rangle$. The orthogonality of the collective modes can be given by
\begin{equation}\label{eq:orthogonality}
\langle\psi_{i}|\sigma_{z}|\psi_{j}\rangle=\int d\boldsymbol{r}\left(u_{i}^{\ast}u_{j}-\upsilon_{i}^{\ast}\upsilon_{j}\right)=\text{sign}\left(\omega_{j}\right)\delta_{ij}.
\end{equation}
According to the Riesz-Schauder theory, the subspace expanded by $|\psi_0\rangle$ is incomplete \cite{Zaanen1964NH, Ida1970some}. To construct a complete eigenspace, one need to introduce the complementary mode $\langle\boldsymbol{r}|\phi_{0}\rangle=(\begin{array}{cc}
\phi_{0}\left(\boldsymbol{r}\right) & \phi_{0}^{\ast}\left(\boldsymbol{r}\right)\end{array})^{T}$ through the equation \cite{Lewenstein1996quantum,
Matsumoto2002quantum, Lundh2002effective, Rammer2007quantum}
\begin{equation}\label{eq:phipsi}
\sigma L_{0}|\phi_{0}\rangle=\alpha|\psi_{0}\rangle,
\end{equation}
where the zero-order Bogoliubov Hamiltonian,
\begin{equation}\label{eq:BogoHam}
L_{0}=\left(\begin{array}{cc}
-\frac{\hbar^{2}}{2M}\nabla^{2}-\mu+V\left(\boldsymbol{r}\right)+2g\left|\psi_{0}\right|^{2} & g\psi_{0}^{2}\\
g\psi_{0}^{\ast2} & -\frac{\hbar^{2}}{2M}\nabla^{2}-\mu+V\left(\boldsymbol{r}\right)+2g\left|\psi_{0}\right|^{2}
\end{array}\right).
\end{equation}
Here the constant $\alpha$ is selected to make
\begin{equation}\label{eq:phi0}
\langle\psi_{j}|\sigma_{z}|\phi_{0}\rangle=\int d\boldsymbol{r}\left(u_{j}^{\ast}\left(\boldsymbol{r}\right)\phi_{0}\left(\boldsymbol{r}\right)-\upsilon_{j}^{\ast}\left(\boldsymbol{r}\right)\phi_{0}^{\ast}\left(\boldsymbol{r}\right)\right)=\delta_{j0}.
\end{equation}
The completeness of the eigenspace is given by \cite{Lewenstein1996quantum,
Matsumoto2002quantum, Lundh2002effective, Rammer2007quantum}
\begin{equation}\label{eq:completeness}
\sum_{j\neq0}\text{sign}\left(\omega_{j}\right)|\psi_{j}\rangle\langle\psi_{j}|+|\phi_{0}\rangle\langle\psi_{0}|+|\psi_{0}\rangle\langle\phi_{0}|=I\sigma_{z},
\end{equation}
where $I$ is the unit operator.

\begin{figure}
\centerline{\includegraphics[width=0.9\columnwidth]{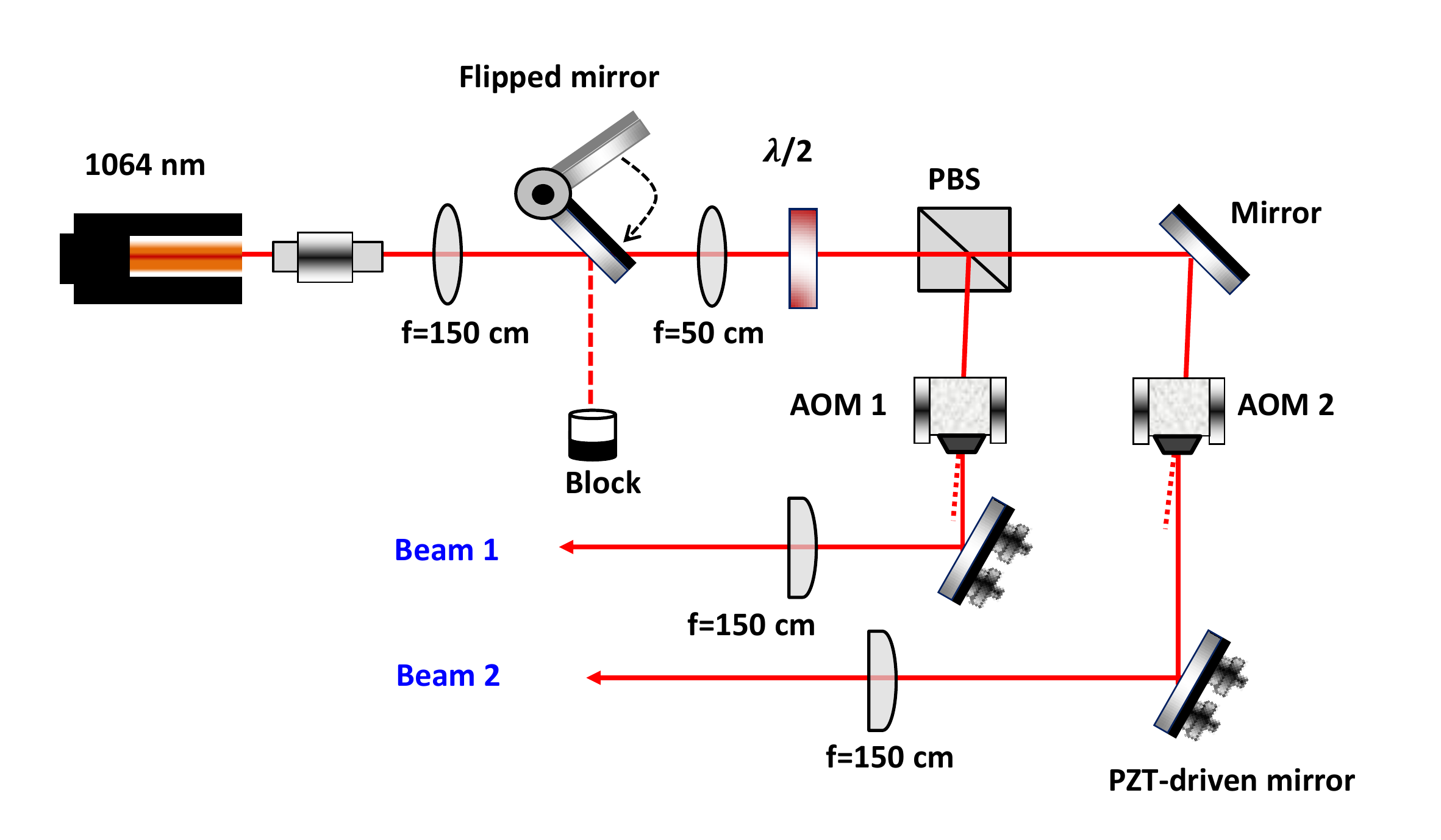}}
\caption{(color online) Schematics for the two optical trap beams of the spherical trap. Two beams come from the one IPG laser with a wavelength of 1064 nm. One acousto-optic modulator (AOM) and one PZT-driven mirror are used together to accurately adjust the position of one beam with a high repeatability. The flipped mirror can reflect the laser beam, making the AOM open most of the time. PBS: polarization beam splitter. $\lambda/2$: half-wave plate}  \label{Figure5}
\end{figure}

\subsection*{Oscillation of the collective modes}
In the above section, we analyze the collective modes in a statistic system. If the system is perturbed by some time-dependent modulation, the collective modes may be excited and the BEC will oscillate with the frequencies of the collective modes. We can measure the oscillation with some observables like the width or the center-of-mass position of the BEC. The corresponding relation between the oscillation forms and excited modes will be discussed in this subsection.

Generally, the oscillating BEC can be described by the normalized wave function
\begin{equation}\label{eq:tExpan}
\psi\left(\boldsymbol{r},t\right)=e^{-i\mu t/\hbar}\left[b_{0}\left(t\right)\psi_{0}\left(\boldsymbol{r}\right)+\sum_{\omega_{j}>0}\left(b_{j}\left(t\right)u_{j}\left(\boldsymbol{r}\right)e^{-i\omega_{j}t}+b_{j}^{\ast}\left(t\right)\upsilon_{j}^{\ast}\left(\boldsymbol{r}\right)e^{i\omega_{j}t}\right)\right],
\end{equation}
where the coefficients satisfy $\left|b_{0}\left(t\right)\right|^{2}+\sum_{j}\left|b_{j}\left(t\right)\right|^{2}=N$. Because all wave functions in the $m=0$ sector are real functions. When $\sum_{j}\left|b_{j}\left(t\right)\right|^{2}$ is small, considering $\psi_{0}$ belongs to the $m=0$ sector, the expectation of any observable $\hat{A}$ that doesn't change magnetic angular momentum can be expanded as
\begin{equation}\label{eq:osc}
\langle\hat{A}\rangle\left(t\right)=\left|b_{0}\left(t\right)\right|^{2}\langle\hat{A}\rangle_{0}+2b_{0}\left(t\right)\left|b_{j}\left(t\right)\right|\sum_{\omega_{j}>0}\left|\langle\hat{A}\rangle_{0j}\right|\cos\left(\omega_{j} t+\theta\right),
\end{equation}
where $\langle\hat{A}\rangle_{0}=\int d\boldsymbol{r}\psi_{0}^{\ast}\left(\boldsymbol{r}\right)\hat{A}\psi_{0}\left(\boldsymbol{r}\right)$, $\langle\hat{A}\rangle_{0j}=\int d\boldsymbol{r}\psi_{0}^{\ast}\left(\boldsymbol{r}\right)\hat{A}\left(u_{j}\left(\boldsymbol{r}\right)+\upsilon_{j}^{\ast}\left(\boldsymbol{r}\right)\right)$, and $\theta$ is a relative phase.

For the monopole mode $j\rightarrow(n,l)=(1,0)$, due to the rotational symmetry of both the ground state and the monopole mode, $\langle x^2 \rangle_{0j}=\langle y^2 \rangle_{0j}=\langle z^2 \rangle_{0j}\neq 0$ and $\langle x \rangle_{0j}=\langle y \rangle_{0j}=\langle z \rangle_{0j}=0$. In general, the variation speeds of the coefficients $b_0(t)$ and $b_j(t)$ are far slower than $\cos(\omega_j t)$. Hence, the excitation of the momopole mode will lead to the periodic oscillation of the BEC width with a frequency $\omega_{mon}\approx\sqrt{5}\omega$. For the quadrupole mode $j\rightarrow(n,l)=(0,2)$ and $m=0$, $\langle x^2 \rangle_{0j}=\langle y^2 \rangle_{0j}=-\langle z^2 \rangle_{0j}\neq 0$ and $\langle x \rangle_{0j}=\langle y \rangle_{0j}=\langle z \rangle_{0j}=0$. For the quadrupole mode, it is out-of-phase between the oscillations in the z direction and x-y plane. While for the mode $j\rightarrow(n,l,m)=(2,1,0)$, $\langle x^2 \rangle_{0j}=\langle y^2 \rangle_{0j}=\langle z^2 \rangle_{0j}=0$, $\langle x \rangle_{0j}=\langle y \rangle_{0j}=0$, but $\langle z \rangle_{0j}\neq0$, only the z-direction center-of-mass (CoM) position oscillates with the frequency $\omega_{(2,1)}\approx\sqrt{19}\omega$ when mode $(2,1)$ is excited. Modes $j\rightarrow(n,l,m)=(2,1,\pm1)$ are not involved in a coherent driving process due to the conservation of magnetic angular momentum. This analysis constructs the base for the fitting of experimental data in the main text.

\subsection*{Calculating the Landau damping rate of the collective mode in the spherical BEC.}

Landau damping, in which low-energy collective mode is absorbed in the transition between thermal excitations, is dominant at finite temperature \cite{Stringari1997PLA, Pitaevskii1997PRA}. We calculate the Landau damping of the collective modes based on the perturbation theory developed by Pitaevskii \textit{et al.} \cite{Gora1998PRL, Stringari1997PLA, Pitaevskii1999PRA}. Accordingly the damping rate is calculated with the expression,
\begin{equation}\label{eq:gamma}
\gamma=\left(\pi/\hbar^{2}\right)\sum_{ik}\left|A_{ik}\right|^{2}\delta\left(\omega_{ik}-\Omega_{osc}\right)\left(f_{i}-f_{k}\right),
\end{equation}
where $f_{\nu}=\left[\exp\left(E_{\nu}/k_{B}T\right)-1\right]^{-1}$  is the thermal occupation of mode $\nu=i,k$ with the temperature $T$ and Boltzmann constant $k_B$, $\Omega_{osc}$ is the eigenfrequency of the oscillation mode, $\omega_{ik}=\omega_{i}-\omega_{k}$ is the frequency difference, and $\delta(.)$ is the Dirac function. Here $A_{ik}= 2g\int d\boldsymbol{r}\psi_{0}[(u_{k}^{\ast}\upsilon_{i}+\upsilon_{k}^{\ast}\upsilon_{i}+u_{k}^{\ast}u_{i})u_{osc}
+(\upsilon_{k}^{\ast}u_{i}+\upsilon_{k}^{\ast}\upsilon_{i}+u_{k}^{\ast}u_{i})\upsilon_{osc}]$  is the transition amplitude of a specific damping channel, where $g$ is the interaction strength, $(\psi_{0}, -\psi_{0}^{\ast})$ and $(u_{\nu},  \upsilon_{\nu})^T$ are the wave functions of the zero-energy Nambu-Goldstone mode and the $\nu$-th collective mode, respectively. The lifetime of the collective mode is given by $\tau=1/\gamma$. In general, due to the uncertainties like the imperfection of the trapping potential and the finite lifetime of the collective mode, the energy level has a finite width. We need to replace the $\delta$-function with a Lorentz distribution with the width $\Delta$, i.e. $\delta(\omega_{ik}-\Omega_{osc})\rightarrow\Delta/(2\pi\hbar)[(\omega_{ik}-\Omega_{osc})^{2}+\Delta^{2}/4]$. In fact, $\gamma$ is unsensitive with $\Delta$ when it is far larger than the average level space and smaller than $\Omega_{osc}$. The calculated lifetime for the monopole and quadrupole modes are about $111.9$ ms and $49.3$ ms, respectively, which show good agreement with the experimental measurements.

\subsection*{Perturbation of gravity potential}
Gravity potential will make the isotropic potential deform and develop odd parity in the z-direction. The odd-parity component of the total potential $\delta V=\lambda z^3$ will couple the monopole mode with the mode $(n,l)=(2,1)$ through processes like the parametric-down conversion of photons in quantum optics. Since the deformation is weak, a perturbation analysis is applicable.

At first, the deformation will change the ground-state wave function and the chemical potential. Under the Thomas-Fermi approximation\cite{Smith2008Cambridge}, the new ground-state wave function $\psi_0^{'}$ and chemical potential $\mu^{'}$ satisfy the equations $g|\psi_{0}^{'}|^{2}=\mu^{'}-V^{'}(\boldsymbol{r})$ and $\int d\boldsymbol{r}|\psi_{0}^{'}|^{2}=N_0$, where $V^{'}=V\left(\boldsymbol{r}\right)+\delta V\left(\boldsymbol{r}\right)$.

Our purpose is to derive the perturbation of the collective-mode wave functions. Since the new ground-state wave function $\psi_{0}^{'}$ can be gauged to be real and positive, the total perturbation term for $L_0$ can be written as $\delta L_{0}\approx(\delta\mu-\delta V)(\begin{array}{cc} 1 & 1\end{array})^{T}(\begin{array}{cc} 1 & 1\end{array})$ when $(\mu^{'}-V^{'}(\boldsymbol{r}))>0$ and $\delta L_{0}\approx-(\delta\mu-\delta V)I_2$ elsewhere. Here $\delta \mu=\mu^{'}-\mu$ and $I_2$ is the two by two identity matrix.  The perturbation expansion of the BdG Eq. (\ref{eq:BogoHam}) can be written as
\begin{equation}\label{eq:ExpBdG}
\left(L_{0}+\delta L_{0}\right)\left(|\psi_{j}^{\left(0\right)}\rangle+|\psi_{j}^{\left(1\right)}\rangle+\cdots\right)=\left(E_{j}^{\left(0\right)}+E_{j}^{\left(1\right)}+\cdots\right)\left(|\psi_{j}^{\left(0\right)}\rangle+|\psi_{j}^{\left(1\right)}\rangle+\cdots\right),
\end{equation}
where $|\psi_{j}^{\left(n\right)}\rangle$ ($|\psi_{j}^{\left(0\right)}\rangle=|\psi_{j}\rangle$) and $E_{j}^{\left(n\right)}$ ($E_{j}^{\left(0\right)}=E_{j}$) are the n-th-order wave function and eigen-energy. The first-three-orders equation are respectively given by
\begin{equation}\label{eq:0thBdG}
\sigma_z L_{0}|\psi_{j}^{\left(0\right)}\rangle=E_{j}^{\left(0\right)}|\psi_{j}^{\left(0\right)}\rangle,
\end{equation}
\begin{equation}\label{eq:1thBdG}
\sigma_z L_{0}|\psi_{j}^{\left(1\right)}\rangle+\sigma_z \delta L_{0}|\psi_{j}^{\left(0\right)}\rangle=E_{j}^{\left(0\right)}|\psi_{j}^{\left(1\right)}\rangle+E_{j}^{\left(1\right)}|\psi_{j}^{\left(0\right)}\rangle,
\end{equation}
and
\begin{equation}\label{eq:2thBdG}
\sigma_z L_{0}|\psi_{j}^{\left(2\right)}\rangle+\sigma_z \delta L_{0}|\psi_{j}^{\left(1\right)}\rangle=E_{j}^{\left(0\right)}|\psi_{j}^{\left(2\right)}\rangle+E_{j}^{\left(1\right)}|\psi_{j}^{\left(1\right)}\rangle+E_{j}^{\left(2\right)}|\psi_{j}^{\left(0\right)}\rangle.
\end{equation}
Integrating the two sides of Eq. (\ref{eq:1thBdG}) with $\langle\psi_{j}^{\left(0\right)}|\sigma_z$ and $\langle\phi_{0}^{\left(0\right)}|=\langle\phi_{0}|$, and noting that $\langle\psi_{j}^{\left(0\right)}|L_{0}|\psi_{j^{'}}^{\left(1\right)}\rangle=E_{j}^{\left(0\right)}\langle\psi_{j}^{\left(0\right)}|\sigma_{z}|\psi_{j^{'}}^{\left(1\right)}\rangle$ and $\langle\phi_{0}^{\left(0\right)}|L_{0}|\psi_{j^{'}}^{\left(1\right)}\rangle=\alpha\langle\psi_{0}^{\left(0\right)}|\sigma_{z}|\psi_{j^{'}}^{\left(1\right)}\rangle$, we can derive
\begin{equation}\label{eq:1thEnergy}
E_{j}^{\left(1\right)}=\text{sign}\left(\omega_{j}\right)\langle\psi_{j}^{\left(0\right)}|\delta L_{0}|\psi_{j}^{\left(0\right)}\rangle,
\end{equation}
and
\begin{equation}\label{eq:1thWF}
\begin{split}
|\psi_{j}^{\left(1\right)}\rangle=&\sum_{j^{'}\neq0,j}\text{sign}\left(E_{j^{'}}^{\left(0\right)}\right)\frac{\langle\psi_{j^{'}}^{\left(0\right)}|\delta L_{0}|\psi_{j}^{\left(0\right)}\rangle}{E_{j}^{\left(0\right)}-E_{j^{'}}^{\left(0\right)}}|\psi_{j^{'}}^{\left(0\right)}\rangle+\frac{\langle\phi_{0}^{\left(0\right)}|\delta L_{0}|\psi_{j}^{\left(0\right)}\rangle}{E_{j}^{\left(0\right)}}|\psi_{0}^{\left(0\right)}\rangle\\
&+\frac{\langle\psi_{0}^{\left(0\right)}|\delta L_{0}|\psi_{j}^{\left(0\right)}\rangle}{E_{j}^{\left(0\right)}}\left(|\phi_{0}^{\left(0\right)}\rangle+\frac{\alpha}{E_{j}^{\left(0\right)}}|\psi_{0}^{\left(0\right)}\rangle\right).
\end{split}
\end{equation}
The condition $\langle\psi_{j}^{\left(0\right)}|\sigma_{z}|\psi_{j}^{\left(1\right)}\rangle=0$ and the completeness relation (\ref{eq:completeness}) have been used in the above derivation. The above first-order perturbation of the wave function $\langle\boldsymbol{r}|\psi_{j}^{\left(1\right)}\rangle=\left(\begin{array}{cc}
u_{j}^{\left(1\right)} & \upsilon_{j}^{\left(1\right)}\end{array}\right)$ has been used in the calculation of $M_{12}$ in the main text. Due to the odd parity of $\delta L_0$,  $E_{j}^{\left(1\right)}=0$. Multiplying the two sides of Eq. (\ref{eq:2thBdG}) with $\langle \psi_{j}^{(0)}|\sigma_z$, we can yield the second-order perturbation of the eigen-energy
\begin{equation}\label{eq:2orderEnergy}
\begin{split}
E_{j}^{\left(2\right)}=&\text{sign}\left(E_{j}^{\left(0\right)}\right)\{\sum_{j^{'}\neq0,j}\text{sign}\left(E_{j^{'}}^{\left(0\right)}\right)\frac{\left|\langle\psi_{j^{'}}^{\left(0\right)}|\delta L_{0}|\psi_{j}^{\left(0\right)}\rangle\right|^{2}}{E_{j}^{\left(0\right)}-E_{j^{'}}^{\left(0\right)}}+\frac{\alpha\left|\langle\psi_{0}^{\left(0\right)}|\delta L_{0}|\psi_{j}^{\left(0\right)}\rangle\right|^{2}}{E_{j}^{\left(0\right)2}}\\
&+\frac{2\text{Re}(\langle\phi_{0}^{\left(0\right)}|\delta L_{0}|\psi_{j}^{\left(0\right)}\rangle\langle\psi_{j}^{\left(0\right)}|\delta L_{0}|\psi_{0}^{\left(0\right)}\rangle)}{E_{j}^{\left(0\right)}}\}.
\end{split}
\end{equation}
Therefore, the perturbed eigen-energy is given by $E_{j}^{'}=\hbar\omega_{j}^{'}=E_{j}+E_{j}^{(2)}, j\neq0$.

\end{widetext}

\end{document}